# A microspectroscopic study of the electronic homogeneity of ordered and disordered $Sr_2FeMoO_6$

Dinesh Topwal, Manju U., Sugata Ray,[*] S. Raj,[†] and D.D Sarma[‡]
Solid State and Structural Chemistry Unit, Indian Institute of Science, Bangalore - 560012, India.

S.R. Krishnakumar[§]
International Centre for theoretical Physics (ICTP), Strada Costiera 11, 34100 Trieste, Italy.

M. Bertolo, S. La Rosa and G. Cautero
Sincrotrone Trieste S.C.p.A., s.s. 14 Km 163, 5 Area Science Park, 34012 Trieste, Italy.

**ABSTRACT**

Besides a drastic reduction in saturation magnetization of disordered $Sr_2FeMoO_6$ compared to highly ordered samples, magnetizations as a function of the temperature for different disordered samples may also show qualitatively different behaviors. We investigate the origin of such diversity by performing spatially resolved photoemission spectroscopy on various disordered samples. Our results establish that extensive electronic inhomogeneity, arising most probably from an underlying chemical inhomogeneity in disordered samples is responsible for the observed magnetic inhomogeneity. It is further pointed out that these inhomogeneities are connected with composition fluctuations of the type $Sr_2Fe_{1+x}Mo_{1-x}O_6$ with Fe-rich ($x>0$) and Mo-rich ($x<0$) regions.

## I. INTRODUCTION

A double perovskite system $Sr_2FeMoO_6$ (SFMO) has come under intense investigations during the last few years, because of its spectacularly large magnetoresistance (MR) property,[1,2] as well as its unusual origin of magnetism.[3] Fully ordered $Sr_2FeMoO_6$ is a half-metallic ferromagnetic system. The B site of the perovskite structure are occupied alternately with localized, nominal $Fe^{3+}$ ($3d^5$, S=5/2) species and highly delocalized, nominal $Mo^{5+}$ ($4d^1$, S=1/2) along all the three axes of the cube in the perfectly ordered double perovskite structure of $Sr_2FeMoO_6$. The localized Fe spin is coupled antiferromagnetically to the spin of the delocalized electron *via* a strong, kinetically-driven mechanism[3] leading to a high Curie temperature for the ferromagnetic ordering of the localized Fe spins with a total moment of 4 $\mu_B$ per formula unit (f.u.). However, the saturation magnetization of most $Sr_2FeMoO_6$ samples reported so far is always found[1,4-7] to be lower than 4 $\mu_B$ per formula unit. This lowering of saturation magnetization is attributed to the presence of mis-site disorders, where some Fe and Mo ions interchange their crystallographic positions, facilitated by their similar ionic radii. Disorder is hence introduced in the system, as the long range Fe-O-Mo-O-Fe ordering of the ideal structure is disrupted. Such disorder necessarily introduces Fe-O-Fe type bonds with antiferromagnetic coupling between them, arising from the strong superexchange interaction between the localized Fe moments. Theoretical calculations[8,9] do suggest that the magnetic moment should decrease due to the presence of disorder. However, the predicted decrease does not match the drastic reduction that is experimentally observed.[6,10] Beside the quantitative aspect of a decrease in the saturation magnetization with increasing disorder, we notice that the magnetization, *M*, as a function of the temperature, $T$ for disordered samples shows qualitative differences between different samples.[10,11] Specifically, in certain cases the M(T) of disordered samples show[10] multiple steps, suggestive of magnetic phase separations that may or may not have an underlying chemical or electronic origin. In order to investigate these important issues, we prepared several highly disordered samples with high concentrations of mis-site defects. We study the spatially resolved electronic structure of such disordered samples using a highly focused beam of photons at a third generation synchrotron source and establish that certain samples with a higher concentration of mis-site defects indeed have signatures of extensive electronic inhomogeneities.

## 2. EXPERIMENTAL DETAILS

Disordered SFMO was synthesized by mixing high purity $SrCO_3$, $Fe_2O_3$, $MoO_3$ and Mo in stoichiometric amounts. The obtained mixture was melt quenched[6] in an inert gas arc furnace. Ordered SFMO sample used for comparison was also made by melt quenching as described above, however, the molten ingot so obtained was further annealed at 1250 $^0$C for 10 hrs in a flow of Ar and $H_2$ (98:2) mixture, and then furnace-cooled to room temperature under the same reducing atmosphere. The samples were characterized using x-ray diffraction and scanning electron microscopy (SEM). Most of the diffraction peaks are same for both ordered and disordered samples, arising from the common underlying perovskite structure. The only difference in the diffraction patterns is the existence of generally weak supercell reflections in case of the ordered sample, arising from the ordering of Fe and Mo sites. The intensities of such supercell reflections are strongly suppressed in the disordered sample,[6] establishing the absence of any long range Fe-Mo ordering. Magnetization measurements showed that the ordered sample has a saturation magnetic moment of 3.2 $\mu_B$ per f.u. at 5 K, while that for disordered samples prepared by us was invariably close to 1 $\mu_B$ per f.u. The grain sizes of all samples synthesized for this study were determined to be about 4-5 $\mu m$ by SEM.

Photoelectron spectroscopy was performed on these samples using photon energy of 95 eV at the spectromicroscopy (3.2L) beam line at the Italian synchrotron light source, Elettra. The light from the synchrotron source is focused on to the sample using a Schwarzschild objective[12] providing a spatial resolution of 0.5 $\mu m$. The sample is moved relative to the photon beam using two perpendicular piezo drivers, allowing one to map out the photoemission spectra from the face of the sample with 0.5 $\mu m$ resolution. The photoemission spectrum is collected using a commercial analyzer with a sixteen channel detector array. All experiments were performed at a base pressure of $1\times10^{-10}$ mbar at room temperature (~298 K). The sample was fractured *in situ* to get a clean surface for the surface sensitive photoemission study in each case.

## 3. RESULTS AND DISCUSSION

We generated intensity images by recording spectral intensities at two different energy regions of the valence band spectrum as shown in Figure 1a. 'Image 1' is formed by adding intensities from the near Fermi edge region, collected by all 16 channels, while 'image 2' is formed by adding intensities from central 8 channels (Channel 5-12) around 5.6 eV which is mainly contributed by O 2$p$ states. In order to avoid any drift between the images, both the images were acquired simultaneously. The observed color contrasts in the images arise from the variation of spectral feature at different parts of the exposed surface. According to the convention used here, the red color denotes the lowest intensity and continuous color variations from red to yellow to green to blue show a gradual increase in intensity. It must be noted that the fracture of such polycrystalline samples to obtain a clean surface for photoemission studies in general does not produce a flat surface. Such uneven fracture produces a sample surface with pronounced topographic features that have strong influence on the spectral intensity due to the variations in the positioning and the orientation of the sample surface locally with respect to the focus point of the electron analyzer; as a consequence additional color contrasts are observed in the raw images ('image 1' and 'image 2'), which can very often be misleading. Therefore, it is preferable to use a ratio image ('image 1'/'image 2'), with the assumption that sample topography affects both the images in a similar way and by division we can mostly get rid of the topographic effects. Hence the color contrast in the ratio images by and large represents the actual variation in the ratio of spectral intensities across the sample surface. Such a division essentially means a normalization of the intensity near the Fermi edge with respect to the corresponding intensity of the O 2$p$ states. Ratio images obtained from disordered and ordered samples are shown in Figure 1b. We see a large variation in color contrast for the disordered sample while the ordered sample appears to be more or less homogeneous, when plotted on the same color scale. Such variations establishing a strong spatial dependence of the intensity of the spectral feature close to the Fermi energy relative to that at the O 2$p$ feature at 5.6 eV binding energy, prove that there is an extensive electronic inhomogeneity in this disordered sample, in contrast to the ordered sample.

To understand the detailed nature of such electronic inhomogeneities observed in the images, we recorded a large number of spatially resolved high resolution spectra on freshly cleaved ordered and disordered samples at various points on the sample surface. We show these spectra in Figure 2a by normalizing all spectra at the peak of the valence band at about 5.6 eV. Spectra from the ordered sample are shown with black solid lines, while those of the disordered sample are shown with red solid lines. This comparative figure makes it evident that the line shape of the valence band spectrum of the ordered sample does not show any significant variation between different points on the sample surface, whereas the disordered sample exhibits wide variations in the spectral shape over the sample surface. This clearly establishes that there is some chemical / electronic inhomogeneity in the disordered sample investigated here. Various spectra obtained on the disordered sample can be broadly classified into three types as shown in Figure 2b. Type-I which has higher spectral weight at 0.3 eV, but has lower intensity at 8 eV, with examples of extreme cases shown by blue closed circles, Type-II which has a strong feature at 8 eV, but low counts at 0.3 eV, represented by red solid circles, and Type-III which is the intermediate case, represented by black solid lines. It can also be seen from Figure 2a that the spectra from the ordered sample is very similar to the Type-III one, obtained from the disordered sample, suggesting that the disordered sample has a wide range of compositional/electronic variation, distributed on both sides of the composition/electronic structure of the ordered sample. These variations of the spectra from the disordered sample can be easily compared with the inhomogeneity suggested by the image shown in Figure 1b for the disordered sample, where red represents a relatively lower intensity (Type-II) and blue a higher intensity (Type-I) in the spectral region near $E_F$.

From band structure calculations and experiments performed at various photon energies on ordered samples, various spectral features in the valence band region have already been identified.[13,14] Based on these earlier works, the nature of inhomogeneity present in this disordered sample can be identified. The spectral region at and around $E_F$ has predominance of Mo 4$d$ $t_{2g}\downarrow$ character. This suggests that the spectra (Type-I) with a higher relative intensity near $E_F$ arises from Mo rich regions, while the spectra with the lower intensity in that energy range (Type-II) originates from Mo deficient regions. From resonant photoemission studies carried out on ordered SFMO samples, it has been concluded[13] that the feature present at 8 eV is correlation driven and has contributions from both Fe and Mo states. The photon energy (95 eV) used for these spectro-microscopy measurements is close to the photon energy which exhibits the Cooper minimum in the photoionization cross-section of localized Mo 4$d$ states.[13] Hence the spectra of Type-I which even though originates from Mo rich regions, have a lower intensity at 8 eV. The photoionization cross section of Fe 3$d$ at 95 eV photon energy is larger compared to that of Mo 4$d$ states, possibly explaining the higher intensity at 8 eV in the Type-II spectra associated with Fe rich regions. In the Fe rich compositions, one expects nearest neighbor Fe-O-Fe interactions replacing some of the Fe-O-Mo bonds in the fully-ordered, stoichiometric $Sr_2FeMoO_6$. Such interactions are expected to influence the low energy Fe 3$d$ significantly in Fe rich regions; this may be responsible for the observed drop in the intensity for Type-II spectrum at 1.5 eV, this region being known[13] to arise from Fe 3$d$ states in $Sr_2FeMoO_6$. While we have discussed the gross differences between the different types of spectra, represented by their extreme cases in Figure 2b, Figure 2b clearly shows that the variation of the spectral shape is smoothly distributed over the entire spectral types. This suggests that the concentrations of Fe and Mo vary continuously over a wide range of composition in this disordered sample. We indeed verified this observation by preparing[15] samples over a wide compositional range $-1 \leq x \leq 0.25$ around the nominal $Sr_2FeMoO_6$ and comparing with various spectral types obtained here.

In order to correlate the spatially resolved spectroscopic results reported here with the magnetic properties of such samples, we note that the magnetization as a function of temperature, $M$ (T), for disordered samples often show large variations from one sample to another, illustrated by comparing some representative results from the published literature in Figure 3. $M$ (T) for the ordered sample adopted from reference 16, shows the expected behavior from a high temperature ferromagnet with a well defined $T_C$; all reported $M$ (T) for highly ordered $Sr_2FeMoO_6$ invariably have essentially the same behavior, shown as plot 1 in Figure 3 for the specific case. In contrast, $M$ (T) for a highly disordered sample has been reported[10] with multiple steps, shown as plot 2 in Figure 3, suggesting that there are several magnetic transitions with distinct and widely different Curie temperatures for this sample. The variation of $T_C$ with composition, $x$, in $Sr_2Fe_{1+x}Mo_{1-x}O_6$ has been recently reported[15] by us. It has been shown that $T_C$ decreases rapidly with decreasing $x$ in Fe deficient ($x < 0$) regime. Therefore, the lower temperature step in $M$ (T), in plot 2 in figure 3 below 200 K suggests substantially Fe deficient regions of reasonably large dimensions in such samples. This is entirely consistent with our finding of abundance of spectral Type-I in our sample. While $T_C$ is relatively insensitive to the composition in the Fe rich ($x > 0$) regime, evidence of Fe-deficient regions implies that the sample must also contain Fe-rich regions in order to conserve the overall average composition of $Sr_2FeMoO_6$ ($x = 0$). Though the magnetization data cannot provide any evidence for the existence of Fe-rich regimes due to the insensitivity of $T_C$ on composition for ($x > 0$) in $Sr_2Fe_{1+x}Mo_{1-x}O_6$, spatially resolved photoemission spectra of the Type-II clearly establish the presence of Fe-rich regions along with Fe-deficient ones in our specific case of disordered sample. It is interesting to note that the magnetization of disordered samples does not necessarily exhibit presence of lower $T_C$, suggesting extensive magnetic, electronic and compositional inhomogeneities over large length-scales. For example, in Figure 3, plot 3 shows a specific $M$ (T) measurement of a highly disordered sample. The presence of disorder is clearly indicated by the substantially lower value of $M$ (T) compared to the ordered sample at the lowest temperature. However, $M$ (T) behavior does not suggest the presence of drastically different $T_C$'s in this sample. An inspection of the $M$ (T) plot close to the $T_C$ reveals that the transition width is slightly larger compared to that for the ordered sample, suggesting a narrow distribution of $T_C$'s and consequently, a narrow distribution of composition and electronic structure in such samples of disordered $Sr_2FeMoO_6$. We have indeed found examples of disordered $Sr_2FeMoO_6$ where spatially resolved photoemission ratio images and spectra recorded at a large number of different spots on the sample surface, shown in Figure 4, essentially establish the sample to be homogeneous at least over a length-scale of ~0.5 $\mu m$, which is the spatial resolution of the technique.

## 4. CONCLUSION

In conclusion, we have investigated the nature of chemical inhomogeneity in the disordered SFMO sample by performing experiments on both ordered and disordered samples using spatially resolved photoemission spectroscopy. From the ratio images and the valence band spectra collected at various points on the sample surface of a disordered sample, it has been shown that extensive chemical inhomogeneities can be present in disordered samples, with distinct Fe-rich and Mo-rich regions. Such compositional fluctuations over more than a micron length-scale lead to extensive electronic inhomogeneity; this is possibly the underlying cause of multiple magnetic phases, suggested by steps in $M$ (T) plots of some disordered samples reported in the literature. In contrast, the results from ordered samples establish it as a homogeneous system. We have also shown it possible to have disordered $Sr_2FeMoO_6$ samples that appear homogenous over comparable length-scales, evidenced by spatially resolved photoemission data as well as M (T) plot.

## ACKNOWLEDGEMENT

We thank ICTP-Elettra Users Programme and Department of Science and Technology for financial support. S. R. K thanks ICTP for a TRIL Fellowship

# REFERENCES

\*   Present address: Tokyo Institute of Technology, Japan.

†   Present address: Department of Physics, Tohoku University, Sendai 9808578, Japan.

‡   Also at: Jawaharlal Nehru Centre for Advanced Scientific Research, Bangalore 560064, India.  Electronic address: sarma@sscu.iisc.ernet.in

§   Present address: Saha Institute of Nuclear Physics, Kolkata 700 064, India

**FIGURE CAPTIONS**

Fig. 1 (a) Intensity images recorded simultaneously at two different places in the valence band region of disordered SFMO. 'Image 1' is recorded near Fermi edge while 'image 2' is recorded at most intense valance band feature. Blue color in the image denotes larger intensity in the valence band, while red color denotes lower intensity. Both these images carry information about the sample topography and variation in spectral feature in 62 x 62 $\mu m^2$ area on the sample. (b) Ratio images formed by dividing 'image 1' by 'image 2', there by removing the contributions from topography to a large extent. Disordered sample show large color contrast compared to ordered sample. Blue denotes Mo rich region, while red represents is Fe rich region, suggesting larger chemical/electronic inhomogeneity in disordered sample while ordered sample is more or less homogeneous.

Fig. 2 (a) Valence band spectra collected at various points on ordered and disordered sample and normalized at the maximum intensity feature (~ 5.6 eV). Black solid line and red solid line represents the spectra from ordered and disordered sample respectively. (b) Valence band spectra obtained from the disordered sample is classified into three types, Type-I spectra (blue solid circles) originating from Mo rich region, Type-II spectra (red solid circles) originating from Fe rich region, while Type III spectra (black solid lines) resembles the spectra obtained from ordered sample. Type-I and Type-II spectra are represented with examples of extreme cases, of the disordered sample.

Fig. 3 Temperature dependence of magnetization for ordered (plot 1) and disordered (plot 2 and plot 3) SFMO samples, taken from reference 16, 10 and 11 respectively. Plot

1 and Plot 2 were measured with an applied filed of 0.5T while plot 3 was measured with a field of 0.1T. For clarity plot 2 and plot 3 is multiplied by 3.

Fig. 4  Ratio image and spectra recorded at different spots on a different (other than the one shown in figure 1) disordered sample.

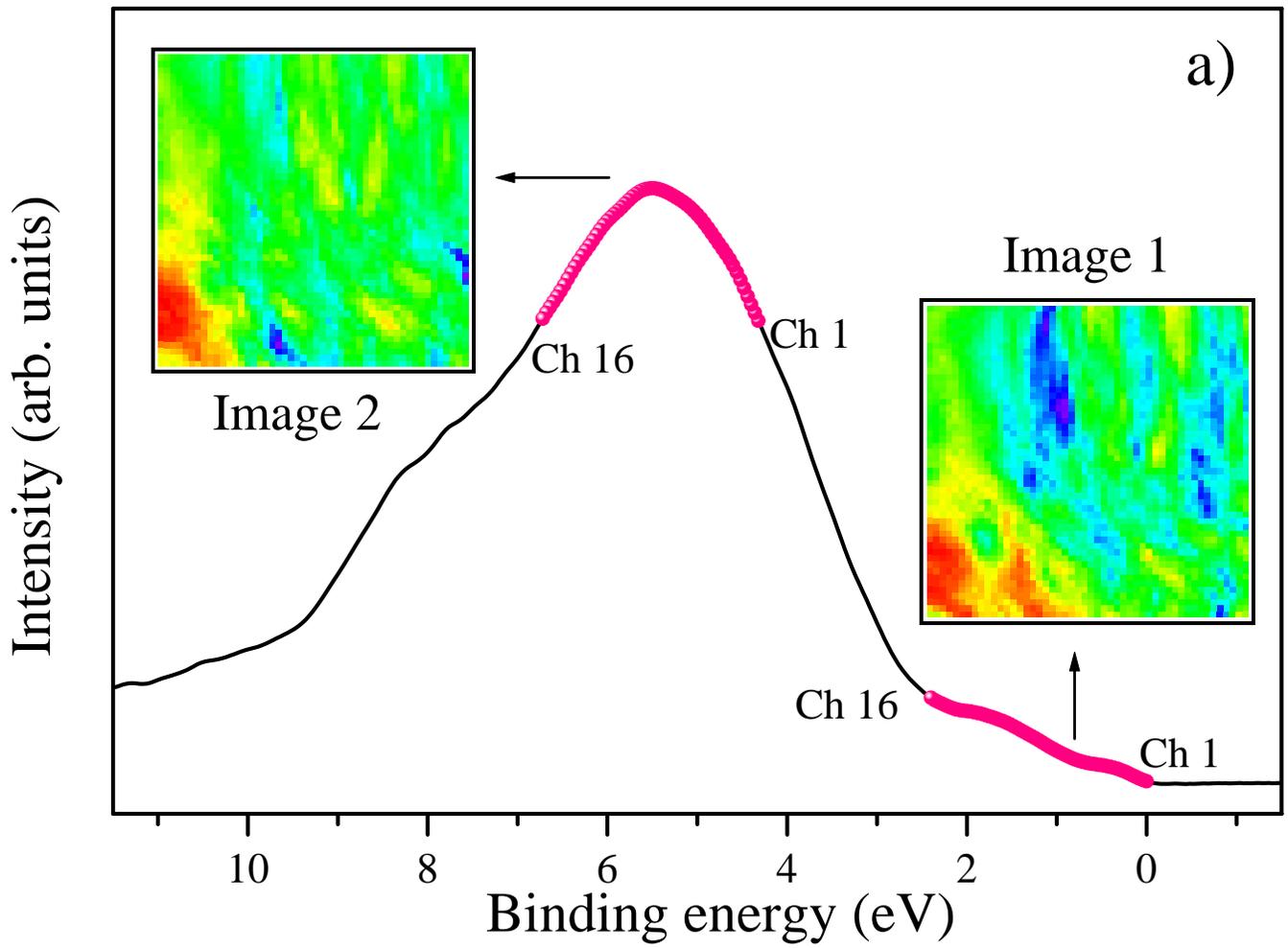

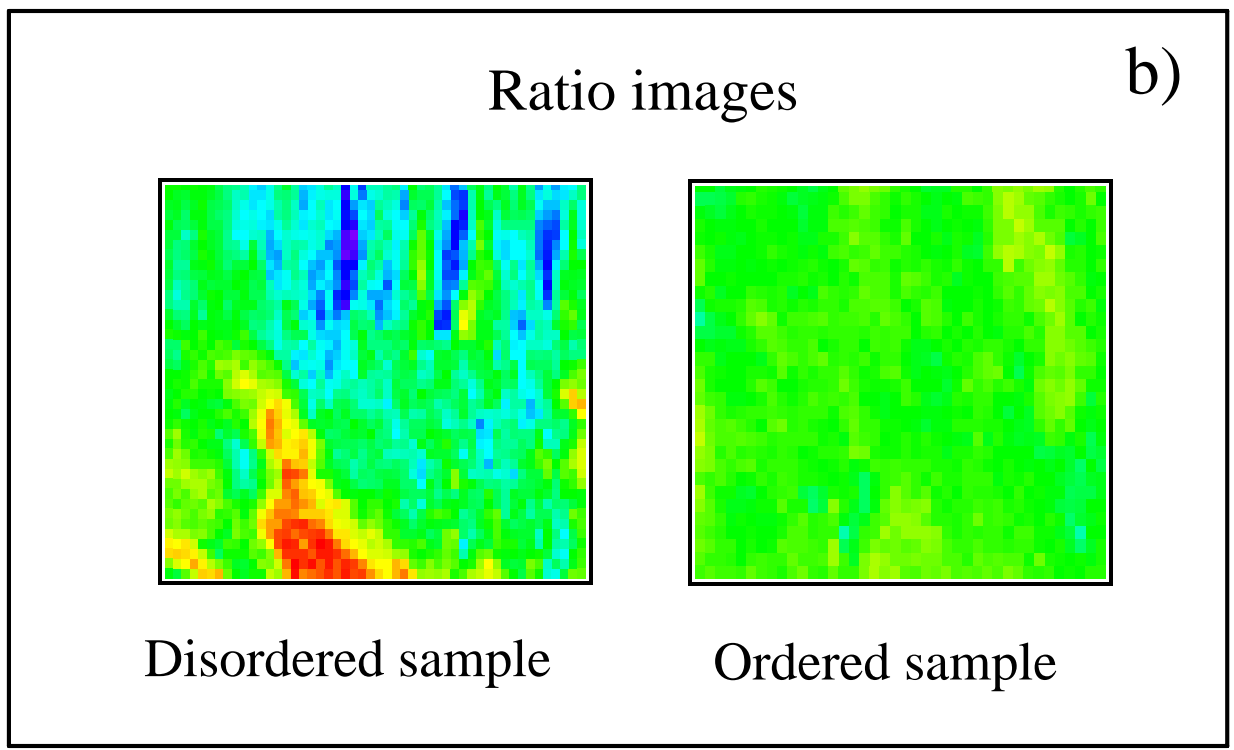

Figure 1, Topwal *et. al.*

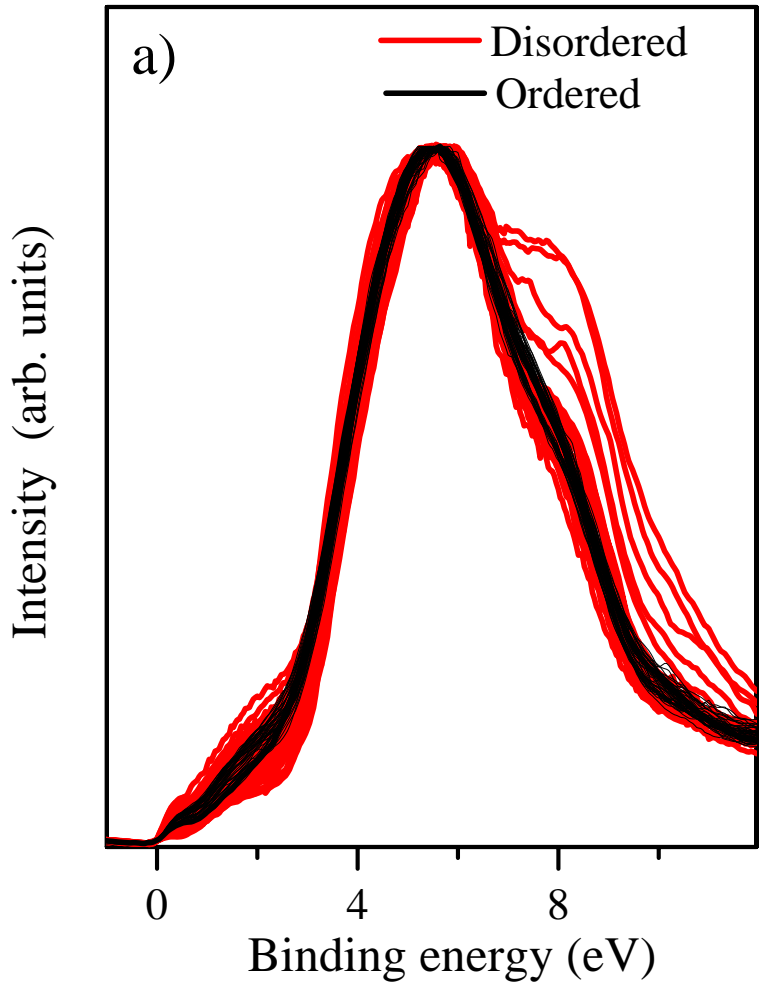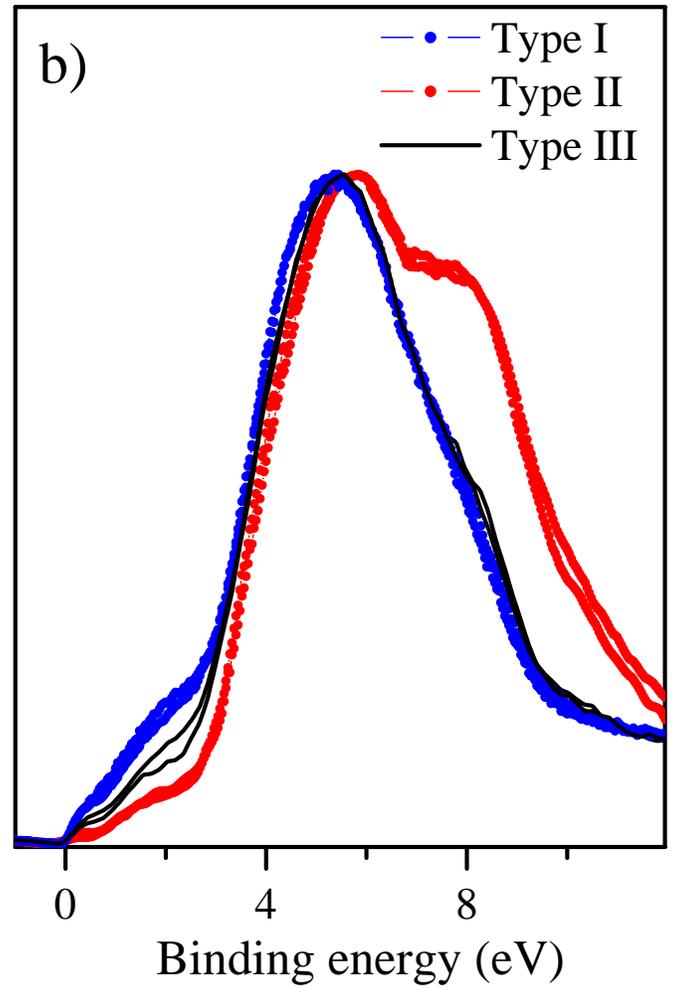

Figure 2, Topwal *et. al.*

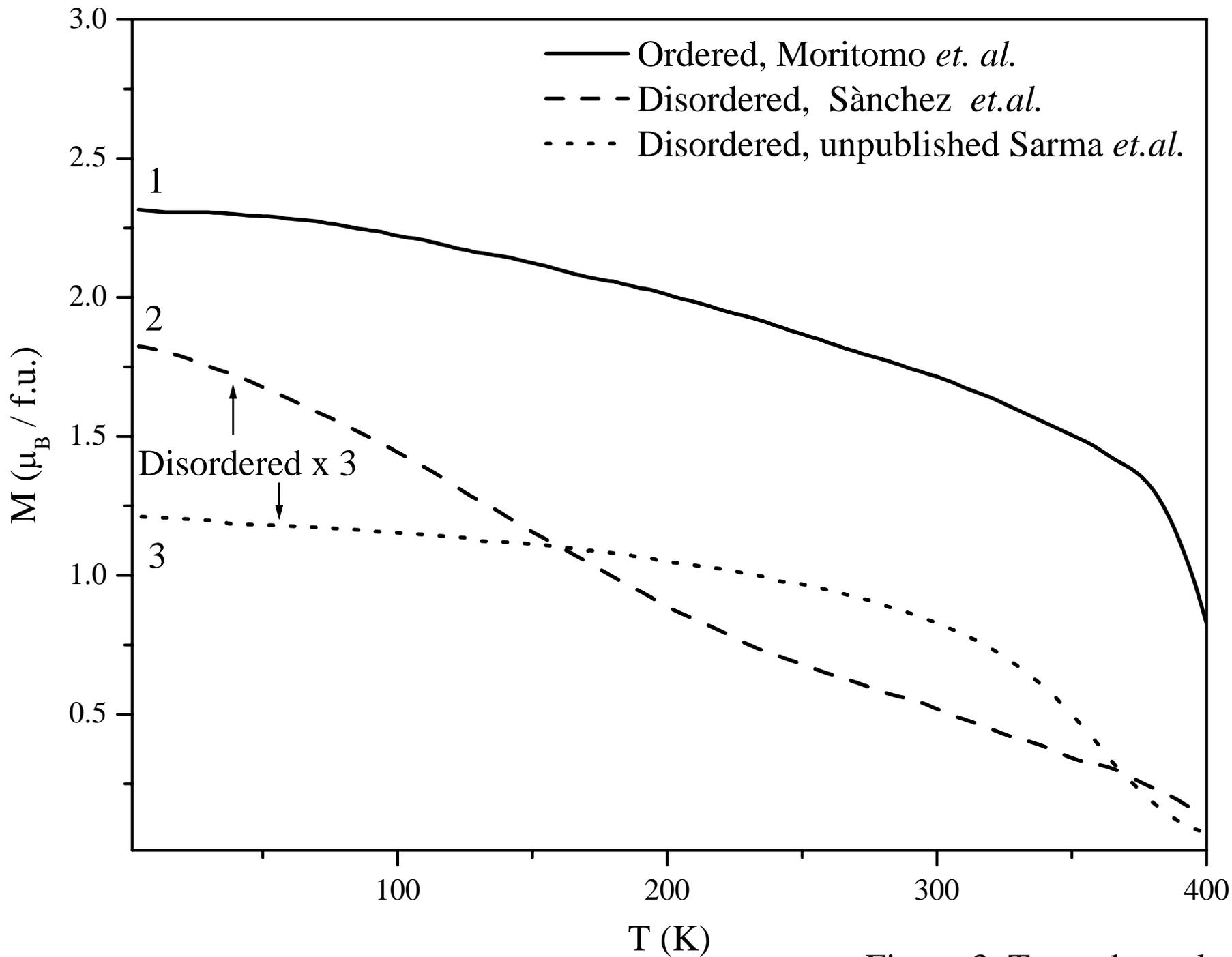

Figure 3, Topwal et. al.

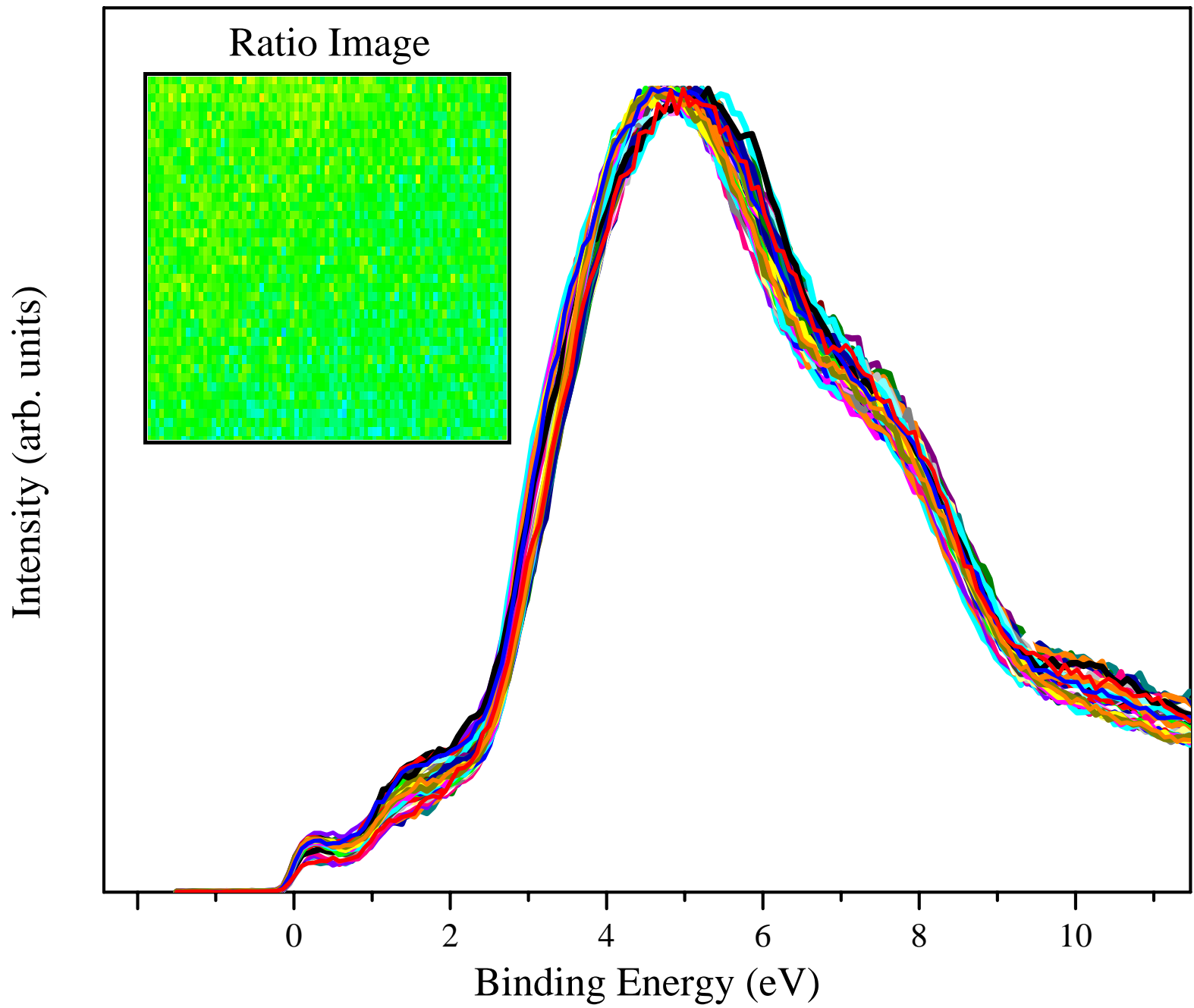

Figure 4, Topwal *et. al.*